# Evaluating Physics Informed Neural Network Performance for Seismic Discrimination Between Earthquakes and Explosions

Qingkai Kong[1*], William R. Walter[1], Ruijia Wang[2], Brandon Schmandt[3]

[1]Lawrence Livermore National Laboratory, Livermore, CA, USA
[2]Southern University of Science and Technology, China
[3]University of New Mexico, USA

[*]Corresponding author: kong11@llnl.gov, Lawrence Livermore National Laboratory 7000 East Ave., Livermore, CA 94550-9234

**Abstract**

Combining physics with machine learning models has advanced the performance of machine learning models in many different applications. In this paper, we evaluate adding a weak physics constraint, i.e., a physics-based empirical relationship, to the loss function (the Physics Informed manner) in local distance explosion discrimination in the hope of improving the generalization capability of the machine learning model. We compare the proposed model to the two-branch model we previously developed, as well as to a pure data-driven model. Unexpectedly, the proposed model did not consistently outperform the pure data-driven model. By varying the level of inconsistency in the training data, we find this approach is modulated by the strength of the physics relationship. This result has important implications for how to best incorporate physical constraints in Machine Learning models.

**Introduction**

The development of machine learning (ML) and its applications show promising results for the automation of many routine tasks in seismology, such as earthquake detection, phase association, event classification and earthquake early warning (Bergen *et al.*, 2019; Kong, Trugman, *et al.*, 2019; Arrowsmith *et al.*, 2022; Mousavi and Beroza, 2022, 2023). So far, the most successful algorithms in many applications use supervised data driven approaches, requiring a ground truth dataset to be provided, and an optimization learning process that iteratively minimizes the estimation errors while searching for a converged model to automatically find the useful features or patterns to address the problem. On the other hand, decades of scientific research on seismological problems have resulted in useful and predictive physical knowledge, such physics-based features that can take the form of algebraic equations, differential equations, empirical relationships, and so on. There are many efforts in various domains to explore different ways to combine physics with machine learning algorithms to take advantage of both areas (Karpatne *et al.*, 2017; Karniadakis *et al.*, 2021; Von Rueden *et al.*, 2021; Cuomo *et al.*, 2022). Figure 1 shows the common approaches to integrate physics knowledge with machine learning models. From a machine learning point of view, we can add physics to the input, use these to modify/design the model structure, or/and include such in the loss function. Central to supervised machine learning algorithms is the need for data to train ML to recognize patterns. One way to incorporate physics is to include physical-law-based simulations in the input training data. To the extent the physical models are correct and match real-world observations, this provides implicit guidance to the ML model to learn at least part of the physics. For example, using more than 6 million different simulated rupture scenarios, Lin *et al.*, (2021) trained a deep learning model to characterize the crustal deformation patterns of large earthquakes in Chilean Subduction Zone, which can successfully estimate the magnitude of 5 real Chilean earthquakes. Similarly, Kong, Inbal, *et al.*, (2019) used a Random Forest model to learn from ground motion model simulation data to estimate magnitude from PGA and distance that were observed on a smartphone seismic network, which yielded decent estimations on test data.

Another option is to encode physics directly into the machine learning model. Many efforts have been tried to modify the machine learning model structure (or design new structure), achieving promising results. For example, a two-branch model has been proposed to take into account physics features with data-driven features that are automatically found by the deep learning model to improve the discrimination of earthquakes and explosions in new regions (Kong *et al.*, 2022). New models have been designed to take advantage of some operations to better extract features that are consistent with physics-based relationships. For instance, the Fourier Neural Operator (FNO) has been designed to learn the integral operator of a differential equation in the Fourier domain, and shows good results in many areas (Li *et al.*, 2020; Kovachki *et al.*, 2021). It has been applied to seismic wave propagation and results are moving towards fast seismic simulations at large-scale (Yang *et al.*, 2021; Li *et al.*, 2022; Song and Wang, 2022a; Yang *et al.*, 2023; Kong and Rodgers, 2023).

Finally, physics constraints can be formulated as a regularization term and added into the loss function of a machine learning model. Since such an approach lets the physics regulate the training of the ML models, it is often referred as physics-informed machine learning or physics-informed neural networks (PINNs), if it is a neural network model (Raissi, Perdikaris and Karniadakis, 2019; Karniadakis *et al.*, 2021; Cuomo *et al.*, 2022). In seismological applications, PINNs have been applied to seismic wave simulations (Karimpouli and Tahmasebi, 2020; Rasht-Behesht *et al.*, 2022; Song and Wang, 2022b; Wei and Fu, 2022), tomography (Chen *et al.*, 2022), and seismic hypocenter inversion (Smith *et al.*, 2021). Many of the PINNs use a differential equation as the physical constraint to the model. In this paper, we adopt a similar strategy, but use an empirical physics relationship as the physical constraint instead of a differential equation.

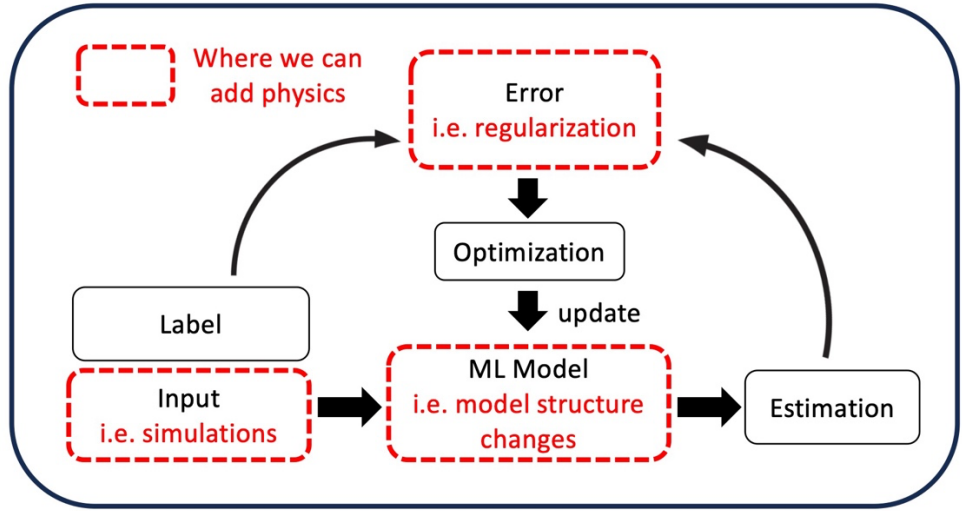

**Figure 1.** The most common approaches to add physics to the machine learning models. Red dashed boxes show where physics is commonly added in from the machine learning point of view.

We choose to evaluate the proposed method in the area of local distance explosion and earthquake discrimination, which is an important area for national security (e.g., Richards and Zavales, 2003; Bowers et al., 2009; Snelson *et al.*, 2013; Koper *et al.*, 2021). Recent studies from the physics-based point of view show that high-frequency P/S amplitude ratios are suitable for small-magnitude event discrimination (O'Rourke, Baker and Sheehan, 2016; Pyle and Walter, 2019, 2021; Wang *et al.*, 2021). Meanwhile, many purely data driven approaches have been proposed for event discrimination and achieved decent results, though questions remain about their robustness for regions and emplacement conditions that are outside of the training data (Linville, Pankow and Draelos, 2019; Tibi *et al.*, 2019; Kim, Lee and You, 2020). In this paper, we report our exploration of adding known empirical physics relationships into the machine learning model as a loss regularization term. We found with this weak physics-based constraint (the empirical P/S ratio relationship) added, this method does not consistently outperform the pure data driven model, nor the two-branch physics plus ML methodology. Lastly, we verified our hypothesis by increasing the strength of the physics-based constraint and show that it improves model performance.

**Methods**

*PINN Model Structures*

We use a standard Convolutional Neural Network, hereinafter CNN (LeCun, Bengio and Hinton, 2015; Goodfellow, Bengio and Courville, 2016) as the back bone of our machine learning model. Our target problem is a single-station-based binary classification (i.e. earthquake versus explosion). As shown in Figure 2, the inputs are 3-component raw waveforms with 50s (seconds) length, which will pass through 3 CNN layers to automatically extract useful features until they are finally flattened into 128 features for the decision layers to make the final decision. We use dropout layers after the CNN layers (dropout rate=0.3) as a regularization to avoid overfitting. Rectified Linear Unit(ReLU) activation functions are used across the whole network except for the last layer, where two neurons with softmax activation functions (Goodfellow, Bengio and Courville, 2016) are employed to estimate the probability of the waveforms being from an earthquake or explosion.

*The Loss Function*

The loss function consists of two parts, i.e. the data driven loss, and the physics loss as shown in equation (1) as well as the red dashed box in Figure 2. Both terms use the standard Sparse Categorical Cross Entropy (SCCE) (Goodfellow, Bengio and Courville, 2016).

$$Total\ loss = \lambda_{DD}\text{SCCE}(y_{true}, \tilde{y}_{ml}) + \lambda_{PHY}\text{SCCE}(y_{p/s}, \tilde{y}_{ml}) \quad (1)$$

where $\lambda_{DD}$ and $\lambda_{PHY}$ are the weights of the data-driven-based loss and physics-based loss. In our test, we fix the $\lambda_{DD}$ at 1, only change the weight of the physics-based loss. $y_{true}$ is the ground truth class, $y_{p/s}$ is the class based on physics empirical relationship (see equation 2 below) and $\tilde{y}_{ml}$ is the class estimated by the machine learning model.

$$y_{p/s} = \begin{cases} Explosion, if\ \frac{P}{S} > h \\ Earthquake, if\ \frac{P}{S} \leq h \end{cases} \quad (2)$$

where h is the P/S amplitude ratio threshold which can separate explosions and earthquakes.

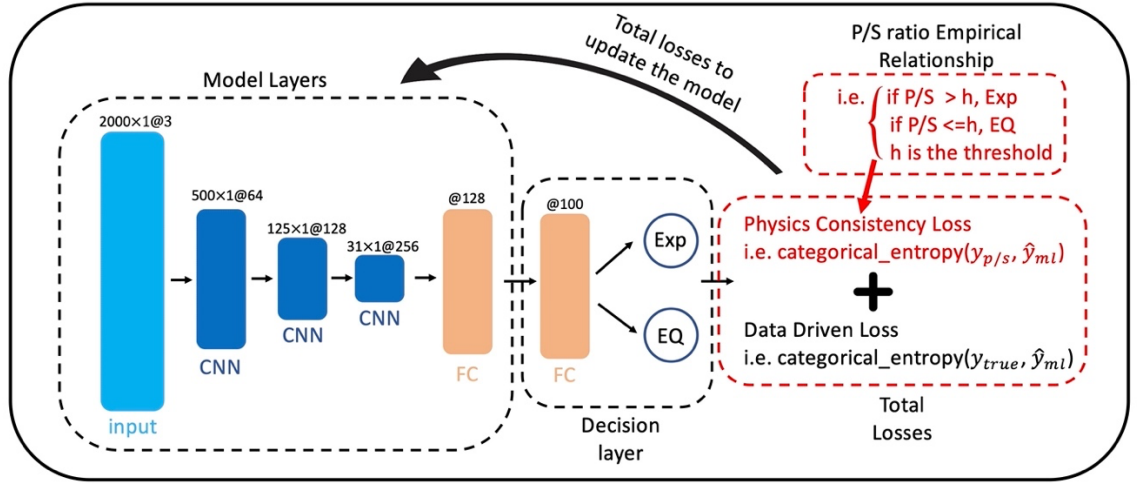

**Figure 2.** The structure of the machine learning model and the added physics consistency loss to regulate the learning of the model. The red texts and boxes show the physics loss we add in addition to the data driven loss. The small text on top of each layer block represents the feature maps in CNN or the number

neurons in Fully Connected Layer (FC), e.g., 500x1@64 represents 64 feature maps with dimensions 500x1.

*P/S Amplitude Ratio*

The P/S amplitude ratios are calculated using the method proposed in (Wang *et al.*, 2021), where the corresponding phases are windowed and filtered between 10 – 18 Hz from the three-component waveforms at all stations. The parameters and regional velocity models used for each region are discussed in more detail in (Wang *et al.*, 2021).

*Training and Testing*

We use the TensorFlow framework (Abadi *et al.*, 2016) for our evaluation. Training, validation, and testing data are described in the data section below. We use Adam as the optimization method (Kingma and Ba, 2017) with an initial learning rate set at 0.001. We run the training with a maximum epoch set at 1,000, and adopt an early stopping mechanism to avoid overfitting, i.e. if the validation accuracy does not change for 30 epochs, the training will be stopped, and the best model will be selected based on the highest accuracy score so far. We use ROC (Receiver Operating Characteristic) curves as well as the Area Under the Curve (AUC) on the testing data as the evaluation metrics (Fawcett, 2006). These metrics can reveal how well one model performs on the testing data compared to other models.

*Other Models*

We compare our proposed approach (Figure 2) with three machine learning models (Figures 3, 4, and 6) described below. Please note all these models have more detailed descriptions in (Kong *et al.*, 2022).

- The model labeled WF+PS is a two-branch model, i.e. one branch is the same convolutional model structure as that shown in Figure 2, which automatically searches for the useful features, and the other branch is a physics-based branch which takes in P/S amplitude ratio as an input, and then combines the features from the two-branch to make the decision.

- The model labeled as WF is the standard convolutional neural network model shown in Figure 2 without the physics-based loss term, this model can be viewed as a purely data driven approach.

- The model labeled PS is a simple ANN (artificial neural network) model to find the best decision boundary for P/S amplitude ratio, which can be viewed as the physics-based feature provided.

**Data**

Here we retain the same four-region dataset used in (Wang *et al.*, 2021; Kong *et al.*, 2022), so we only give a brief summary, as more details can be found in those papers. This dataset also has been released as a subset of an eight-region dataset (Maguire et al., 2024). This compiled dataset includes local distance seismic data from (1) Source Physics Experiment (SPE) Phase I in Nevada (Snelson *et al.*, 2013), (2) Bighorn Arch Seismic Experiment (BASE) (Yeck *et al.*, 2014; Worthington *et al.*, 2016) in Wyoming, (3) Imaging Magma Under Mount St. Helens project (MSH) in Washington (Kiser et al., 2016; Ulberg et al., 2020) and (4) The Salton Seismic Imaging Project (SSIP) in Southern California (Han *et al.*, 2016; Fuis *et al.*, 2017). We limit the distance of recordings to less than 250 km. The same pre-processing steps are used as in (Kong *et al.*, 2022), which includes removing the mean and trend, applying a Hanning window, band-pass filtering from 1 – 20 Hz, re-sampling to 40 Hz, and cutting each waveform into 2,000 data points (50 s) with a random start window (0-5 s) before the origin time. To ensure the dataset is balanced and shift-invariant we randomly select the start time five times for each earthquake record, and 21 times for each explosion record; this data augmentation process leads to a total of 173,385 and 178,059 records for the earthquakes and explosive events, respectively, which are roughly comparable for training and testing purposes. The details of the data, such as location maps, raw data and augmented data distance, magnitude, depth and P/S ratio distributions are shown in Figures S1, S2 and S3 (Kong *et al.*, 2022).

**Results**

We evaluate the performance of the proposed model on its generalization capability by applying the trained model to new region (that was never present in the training data), in the hope that the newly developed model with additional physics constraint will perform better. Figure 3 shows an example when we use P/S amplitude threshold 1 (i.e. h=1 in equation 2) and weight of the physics-based loss at 1 (i.e. $\lambda_{PHY} = 1$ in equation 1). The model is trained using data from three regions while the tested on the fourth region (shown as the title for each panel in Figure 3). After looping through the four scenarios, unexpectedly, the PINN approach only shows better performance for MSH, when compared to the data-driven model alone (the green WF model); for BASE and SSIP tests, the two metrics achieved comparable accuracies and recall (all within the bounds of uncertainties). When the SPE region is used as the test dataset, our PINN model even shows degraded performance. None of the PINN tests outperforms the two-branch model (blue WF+PS model) that added the P/S amplitudes directly into the model (Kong et al., 2022).

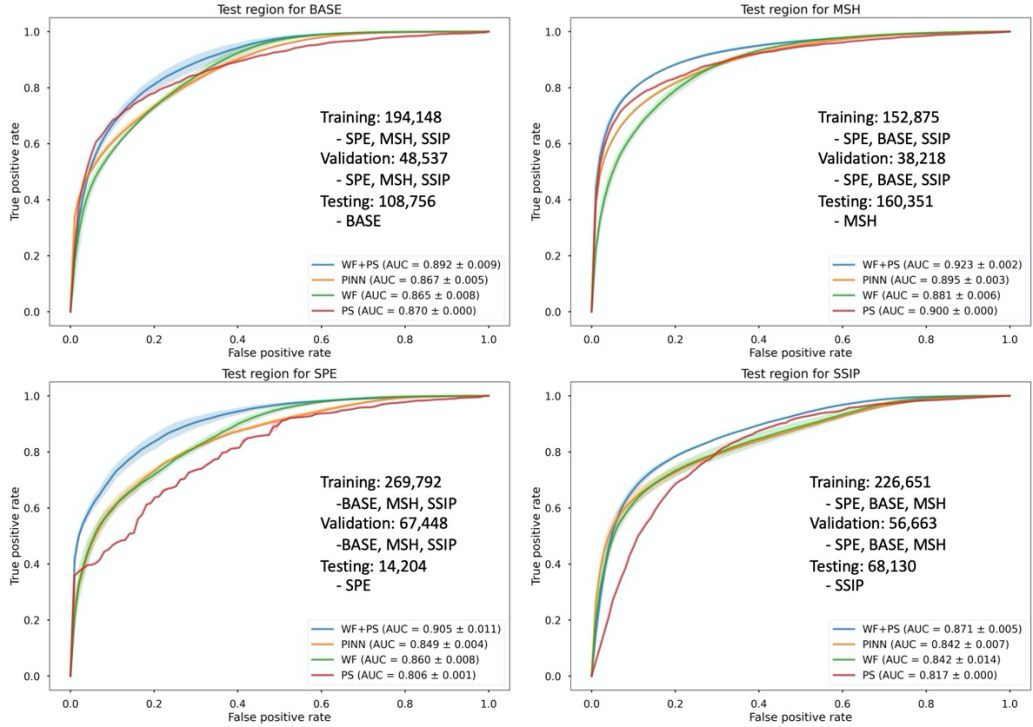

**Figure 3.** Example of classification performance metrics. The ROC curve using training data from any of the three regions and testing on the new fourth region for five random initializations with mean (solid lines) and standard deviation (shaded areas). Number of training, validation and testing data are shown in each panel along with the regions the data come from. The red curves are the designed physics-informed approach. The blue curves show the two-branch model with deep learning and physics parameters branches. The orange, green and red curves are the PINN model proposed, and models only with the deep learning or physics parameters branch. The AUC is shown in the legend.

To show the effect of different levels of physics-based constraints, we conduct experiments with different $\lambda_{PHY}$ values, such as 0.1, 0.5, 1, and 2. The average AUCs for multiple runs for different region tests are shown in Figure 4. The two-branch model is consistently the best model for each region. There is a general trend that when the physics alone model (the PS model) has better performance compared to the data-driven model (WF), the PINN model also shows better performance, such as in the test regions BASE and MSH. For regions like SPE and SSIP, the PS model performed worse than the WF, which leads to inferior performance of the PINN model. Depending on the accuracy of the physics-based model, different levels of physics-based constraints are needed to achieve the best results, which makes the $\lambda_{PHY}$ hyperparameter region dependent.

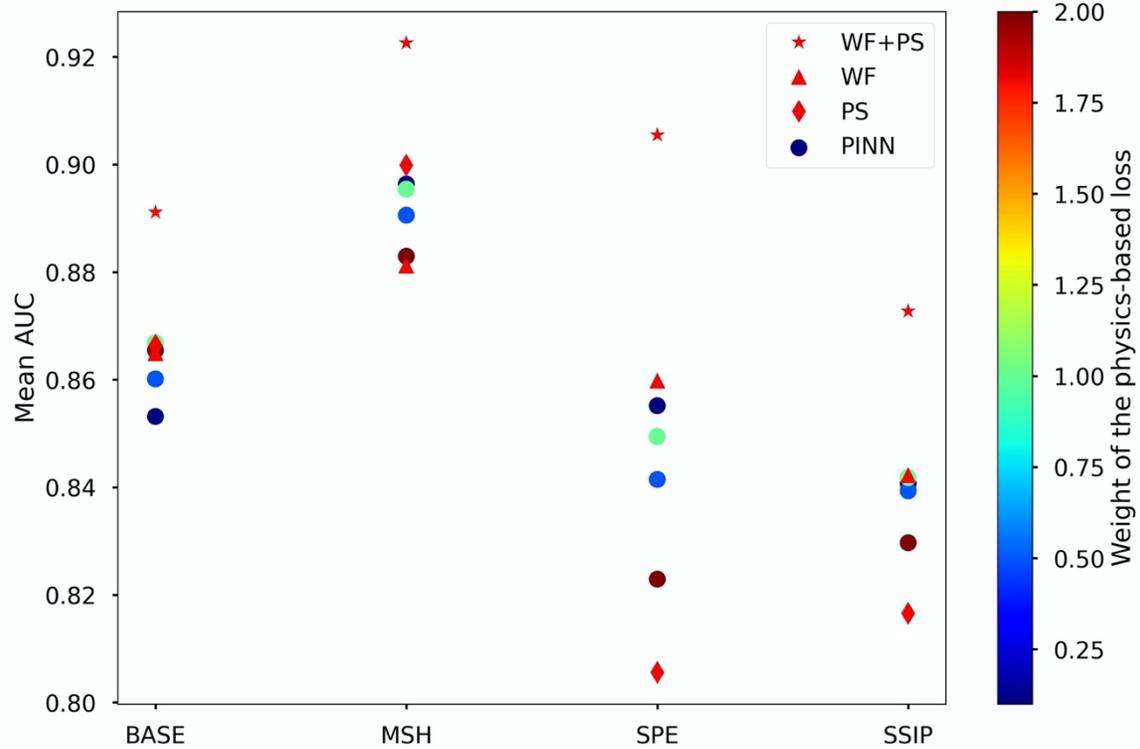

**Figure 4.** Averaged AUC from 5 different runs for each test region for different models. The colors of the circles show the different physics-weight parameter values used in equation 2, i.e., 0.1, 0.5, 1 and 2.

As we can see from the previous section, the proposed PINN approach by integrating the P/S amplitude ratio empirical relationship does not show consistent improvement over the pure data-driven model in terms of testing the performance on a new region. We conducted many different tests, such as using different P/S ratio thresholds (i.e. different h's in equation 2), but none of these tests show improvement over the pure data-driven approach. We speculate such underperformances are associated with the limitations of the physical model: there are samples in the data that cannot be clearly separate based solely on the single-station P/S ratio. For example, (Zhang, 2023) reports the explosive and earthquake sources cannot be discriminated at certain azimuths using P/S ratios due to the radiation patterns, and (Pyle and Walter, 2019, 2021) also show the similar observations that the network averaged P/S ratios may be needed to reliably discriminate the explosions. Figure 5a shows the normalized cumulative distribution of the P/S ratios for the earthquakes and explosions with a threshold of 1: about 10% of earthquake samples have P/S ratios that are inconsistent with the classification threshold (i.e., P/S ratio > 1), while about 35% of explosion samples exhibit inconsistency with the classification threshold (i.e., P/S < 1). Consequently, about 24% of the total data have inaccurate labels. To test this hypothesis, we remove the mis-labeled samples for both the earthquake and explosion data, and form a cleaned empirical physics relationship, as shown in Figure 5b.

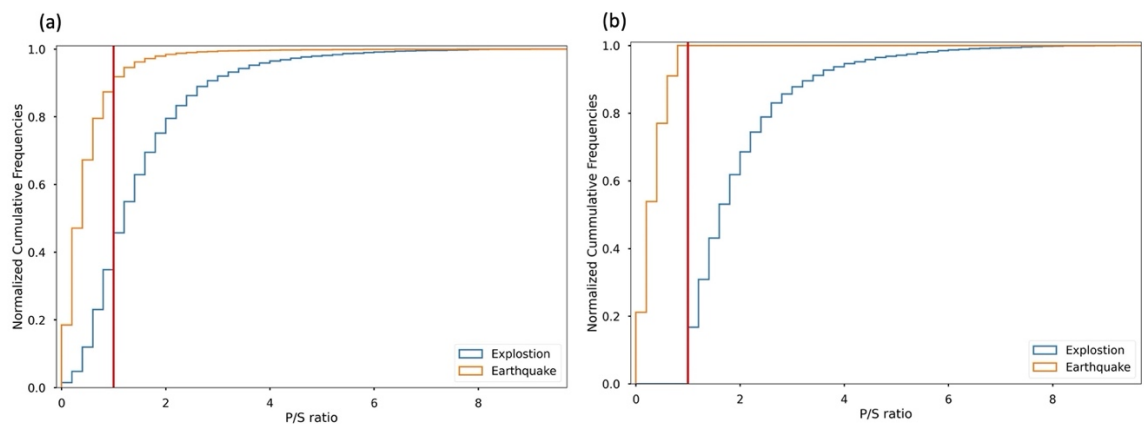

**Figure 5.** The cumulative frequencies of the P/S ratios for the dataset before and after removing the inconsistently labeled samples. (a) The P/S ratio distribution for the dataset used. (b) The P/S ratio distribution after removing the samples inconsistent with a threshold at 1. The vertical red line shows the threshold of P/S ratio at 1.

With this upgraded dataset, we redo the new region tests, and show the results in Figure 6. We see in all 4 region tests, the PINN models are consistently performing much better than the pure data-driven models (WFs), except for SSIP where the improvement is only slight (still within the uncertainties). This confirms our speculation that the accuracy of empirical physics significantly affects the effectiveness of the regularization from the physics-based loss for our proposed method. In many of the previous PINN studies (Smith *et al.*, 2021; Chen *et al.*, 2022; Rasht-Behesht *et al.*, 2022), a differential equation is usually the vehicle where the physics resides, serving as a much stronger version of physics compared to the empirical physics relationship we use here.

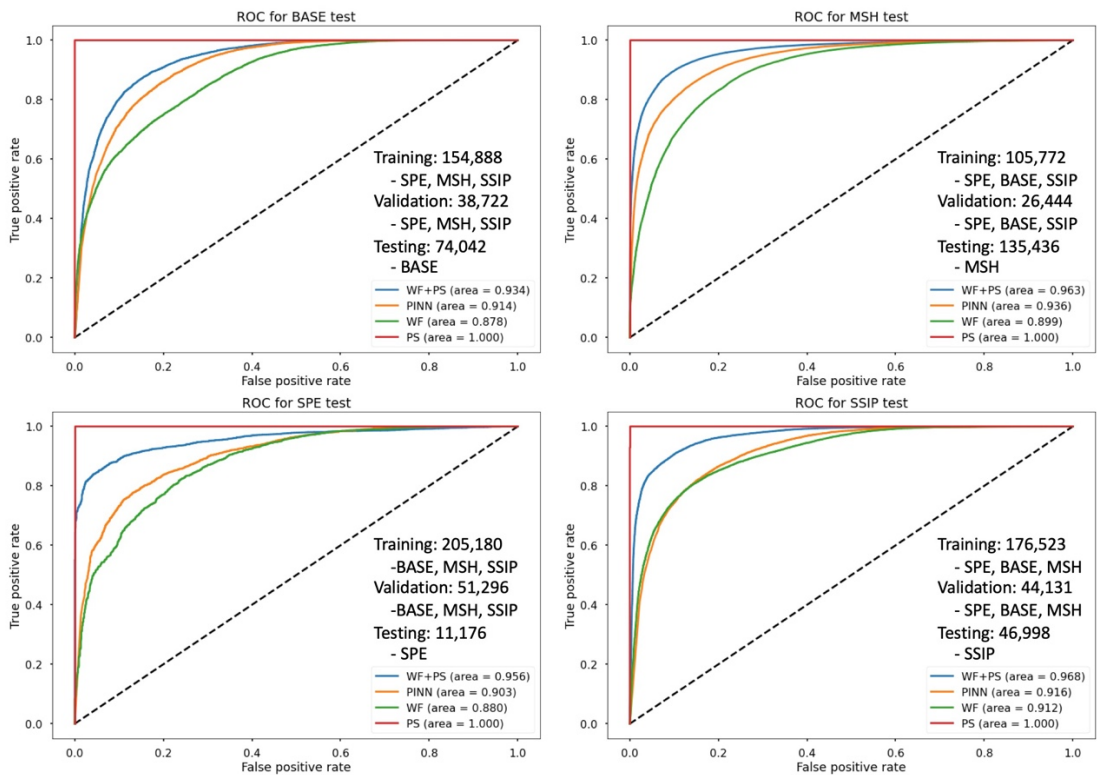

**Figure 6.** The performance metrics after removing the inconsistent samples. For legend or details of the figure, please refer to figure 3.

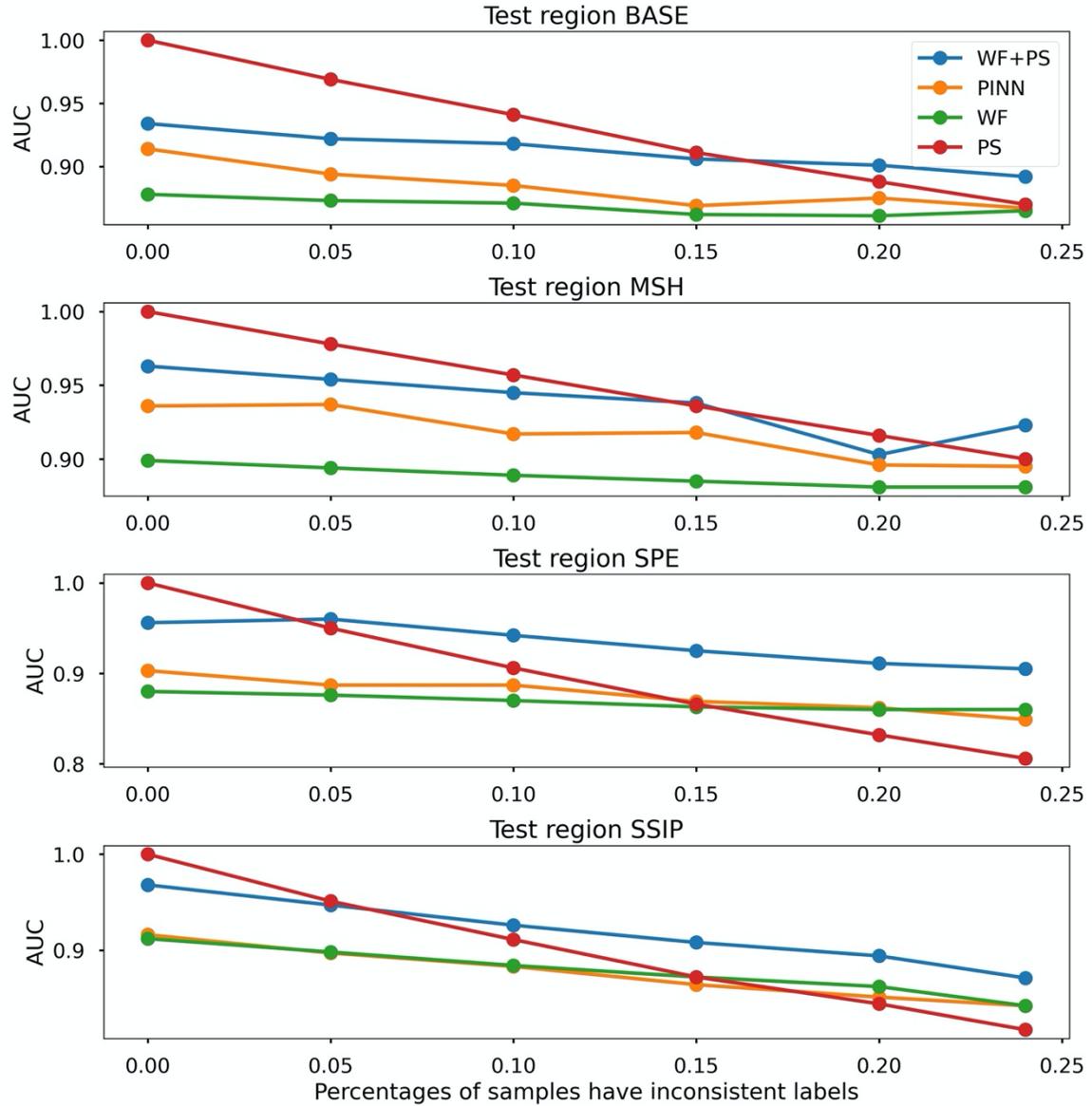

**Figure 7.** The performance of the models with different percentages of samples that have inconsistent labels. Different colors represent the different type of models. For legend of the figure, please refer to Figure 3.

To better understand the effect of the different levels of physics-based constraint accuracy, we conduct experiments by varying the percentage of the inconsistent labels within the dataset, from 0, 5%, 10%, 15%, 20%, and 24% as shown in Figure 7. As expected, the gaps between the PINN and the WF models decrease with the increase of the percentage of the inconsistent samples in the data, i.e. with weaker physics constraints, some regions show a reversal in the relative performance of the PINN and WF models, such as for SPE and SSIP test regions, which cross at 20% and 10%, respectively.

**Discussion and Conclusion**

In this paper, we evaluate the PINN approach using an empirical relationship in the earthquake and explosion discrimination application. We show this approach performs and generalizes (i.e., apply the trained model on the new region) better than a pure deep learning model, on the condition of sufficiently accurate empirical physics constraints. We demonstrate that additional generic physics-based information could help the machine learning model to generalize better, but special attention should be paid to the quality of such physics-based information when incorporating it.

Comparing to the two-branch model which processes the physics-based information directly in one of the branches, the PINN approach never outperforms it. We think the difference in performance is due to explicitly adding the physics information in the two-branch model, so such information is used in both the training and inferencing stages. Thus, when applying this model in a new region, the physics information (P/S ratio) is needed for this new region as well. In contrast, the PINN approach only adds in the physics information implicitly in the loss function to guide the training, not contributing at the inferencing stage. In some sense this leads to less effective overall training of the model than in the two-branch case, combined with demonstrating there is additional utility for the P/S values, despite their level of inconsistency, when used in the inferencing stage. While the PINN approach is easier to set up and use, we have shown that for typical and noisy real world seismic data the two-branch model can consistently outperform the other models.

During our evaluation tests, we also conducted a few more tests with variable ways adding physics constraints, but none of these achieved significant improvements. First, on the original dataset (in which the P/S ratio relationship has about 24% inconsistently labeled samples), we tried the optimal P/S amplitude ratio for each region (the threshold yields the least error in terms of separating the two classes), but the results show a similar trend, not a consistently better performance for the PINN approach. Second, instead of directly adding the physics error into the loss function, we used a sequential training approach (Amini, Haghighat and Juanes, 2023) that we first trained the deep learning model with the regular data-driven loss, and then subsequently trained the model with the physics-loss, but we did not observe any improvement. Third, we also tried to use the physics-loss on the two-branch model by both explicitly and implicitly adding the physics constraints: the performance remains comparable with the two-branch model, even degraded for a few cases. Thus, we conclude that the physics constraints added explicitly in the two-branch model is enough to capture the physics, adding more constraints through the error term can potentially make the training more complicated and degrade performance.

Even though the PINN approach proposed in this paper does not promise practical applications, we think it provides additional insights into how one should add and design a physics involved machine learning model to effectively improve their generalization capability, as well as what factors control the performance. In particular, incorporating physically transportable but imperfect constraints through two-branch model appears to better balance the data driven training and help prevent ML over-training on features that may not be transportable (such as path differences rather than source differences). Our findings and suggestions have important implications on how to best incorporate physical constraints in Machine Learning models.

**Data and Resources**

**Acknowledgements**

This Source Physics Experiment (SPE) research was funded by the National Nuclear Security Administration, Defense Nuclear Nonproliferation Research and Development (NNSA DNN R&D). The authors acknowledge important interdisciplinary collaboration with scientists and engineers from LANL, LLNL, NNSS, and SNL. The views expressed in the article do not necessarily represent the views of the U.S. Department of Energy or the U.S. Government. This research was performed in part under the auspices of the U.S. Department of Energy by the LLNL under Contract Number DE-AC52-07NA27344. This is LLNL Contribution LLNL-JRNL-861256. B.S. acknowledges research support from AFRL FA9453-21-02-0024. We also thank the researchers of the datasets used in this study, including Source Physics Experiments (SPE), Bighorn Arch Seismic Experiment (BASE), Salton Seismic Imaging Project (SSIP), and imaging magma under Mount St. Helens (iMush; i.e., MSH). EarthScope Data Services provides open access to all the data used in the study. EarthScope seismic data facilities are supported by the National Science Foundation under Cooperative Support Agreement EAR-1851048. We appreciate useful discussions with Jiun-Ting (Tim) Lin and Steve Myers at the Lawrence Livermore National Laboratory. All the analysis was done in Python using the TensorFlow deep learning framework (Abadi *et al.*, 2016) and seismological analysis tools from Obspy (Beyreuther *et al.*, 2010; Krischer *et al.*, 2015), we thank the awesome Python communities for making everything openly available.

**Competing interests**

The authors acknowledge there are no conflicts of interest recorded.